\documentclass[prl,twocolumn,%showpacs,
tightenlines,superscriptaddress,preprintnumbers,floatfix]{revtex4}

\def\lsim{\raise0.3ex\hbox{$<$\kern-0.75em\raise-1.1ex\hbox{$\sim$}}}
\def\gsim{\raise0.3ex\hbox{$>$\kern-0.75em\raise-1.1ex\hbox{$\sim$}}}

\usepackage{graphicx}

\newcommand{\beq}{\begin{equation}}
\newcommand{\eeq}{\end{equation}}
\newcommand{\bqa}{\begin{eqnarray}}
\newcommand{\eqa}{\end{eqnarray}}

\begin{document}

\title{Viscosity Information from Relativistic Nuclear Collisions: How Perfect is the Fluid Observed at RHIC?}

\preprint{INT PUB 07-14}

\author{Paul Romatschke}
\affiliation{Institute for Nuclear Theory, University of Washington,
Box 351550, Seattle WA, 98195, USA}
\author{Ulrike Romatschke}
\affiliation{Department of Atmospheric Sciences, University of Washington,
Box 351640, Seattle WA, 98195, USA}
\date{\today}

\begin{abstract} 
Relativistic viscous hydrodynamic fits to RHIC 
data %from the Relativistic Heavy-Ion Collider (RHIC) 
on the centrality 
dependence of multiplicity, transverse and elliptic
flow for $\sqrt{s}=200$ GeV Au+Au collisions are presented.
For standard (Glauber-type) initial conditions, while data on
the integrated elliptic flow coefficient 
$v_2$ is consistent with a ratio of viscosity over entropy
density up to $\eta/s\simeq 0.16$, data on minimum bias $v_2$ seems 
to favor a much smaller viscosity over entropy ratio,
below the bound from the AdS/CFT conjecture. 
Some caveats on this result are discussed.
\end{abstract}

\maketitle

%\section{Introduction}

The success of ideal hydrodynamics for the description
of heavy-ion collisions at the Relativistic Heavy-Ion Collider
(RHIC)
%\footnote{The Relativistic
%Heavy Ion Collider at Brookhaven National Laboratory} 
has led to the idea of a quark-gluon plasma behaving as a 
``perfect liquid'', with a very small ratio of viscosity
over entropy density 
\cite{Teaney:2000cw,Huovinen:2001cy,Hirano:2002ds,
Tannenbaum:2006ch}.
An answer to the question ``How perfect is the fluid observed at RHIC?'' 
can, however, not be found using ideal hydrodynamics,
but must involve a controlled quantitative understanding
of non-idealities, e.g. viscous effects.
If hydrodynamics can be applied to RHIC physics,
then relativistic viscous hydrodynamics 
should be able to provide such an understanding.
In particular, if one has control over the initial conditions,
it should be possible to determine the size of 
various hydrodynamic transport coefficients, such as the shear
viscosity, by a best fit of viscous hydrodynamics (VH) to 
experimental data. In this Letter, we aim to take a step
in this direction.

For RHIC physics, since particle number in the 
quark-gluon plasma is ill-defined,
the relevant dimensionless parameter for VH %viscous hydrodynamics
is the ratio shear viscosity $\eta$ over entropy density $s$.
Based on the correspondence between Anti-de-Sitter (AdS) space
and conformal field theory (CFT), it has been conjectured \cite{Kovtun:2004de}
that
all relativistic quantum field theories at finite temperature
and zero chemical potential have $\eta/s \ge \frac{1}{4\pi}$. 
To date, no physical system violating this bound has been found.

Neglecting effects from bulk viscosity and heat conductivity,
the energy momentum tensor for relativistic hydrodynamics
in the presence of shear viscosity is  
\beq
T^{\mu\nu}=(\epsilon+p)u^\mu u^\nu-p g^{\mu \nu}+\Pi^{\mu \nu}.
\label{EMT}
\eeq
In Eq.~(\ref{EMT}), $\epsilon$ and $p$ denote the energy density and pressure,
respectively, and $u^\mu$ is the fluid 4-velocity which obeys
$g_{\mu \nu} u^\mu u^\nu \!=\!1$ when contracted with the metric $g_{\mu \nu}$.
The shear tensor $\Pi^{\mu \nu}$ is symmetric, traceless ($\Pi^\mu_\mu\!=\!0$),
and orthogonal to the fluid velocity, $u_\mu \Pi^{\mu \nu}\!=\!0$.
Conservation of the energy momentum tensor and equation of state 
provide five equations for the evolution of the 10 independent components
of $\epsilon,p,u^\mu,\Pi^{\mu\nu}$. The remaining five equations for 
the evolution of $\Pi^{\mu \nu}$ are not unambiguously agreed on
at present \cite{Muronga:2003ta,Heinz:2005bw,Baier:2006um,
Tsumura:2006hn,Koide:2006ef}.
%
%
%
%At present, 
%a matter of complication is that 
%various groups use slightly different
%there is
%still disagreement on the 
%forms of the equations for viscous 
%hydrodynamics
%Neglecting effects from bulk viscosity and heat conductivity,
The results in this work will be based on using the set of equations
%we will use the set of equations 
\bqa
(\epsilon+p)D u^\mu&=&\nabla^\mu p-
\Delta^\mu_\alpha d_\beta \Pi^{\alpha \beta}\, ,
\nonumber\\
D \epsilon &=& - (\epsilon+p) \nabla_\mu u^\mu+\frac{1}{2}\Pi^{\mu \nu}
\langle\nabla_\nu u_\mu\rangle\, ,
\nonumber\\
\Delta^\mu_\alpha \Delta^\nu_\beta D \Pi^{\alpha \beta}
&=&-\frac{\Pi^{\mu \nu}}{\tau_{\Pi}} +\frac{\eta}{ \tau_{\Pi}}
\langle\nabla^\mu u^\nu\rangle
- 2 \Pi^{\alpha (\mu}\omega^{\nu)}_{\ \alpha}
\nonumber\\
&&+\frac{1}{2} \Pi^{\mu \nu}\left[5 D \ln T- \nabla_\alpha u^\alpha
\right],
\label{baseq}
\eqa
%for the evolution of the energy density $\epsilon$, fluid velocity
%$u^\mu$ and shear tensor $\Pi^{\mu \nu}$
%in the energy momentum tensor
where $d_\alpha$ is the covariant derivative, used
%which is used 
to construct the time-like and space-like
derivatives $D\!\equiv\! u^\alpha d_\alpha$ and 
$\nabla_\mu\!\equiv\! \Delta_\mu^\alpha d_\alpha$. %, respectively.
The remaining definitions are 
$\Delta^{\mu \nu}\!=\!g^{\mu \nu}\!-\!u^\mu u^\nu$, 
$\langle\nabla^\mu u^\nu\rangle\!=\!
\nabla^\mu u^\nu\!+\!\nabla^\nu u^\mu\!-\!\frac{2}{3} \Delta^{\mu \nu} \nabla_\alpha
u^\alpha$ and the vorticity 
$\omega_{\mu \nu}\!=\!\nabla_\nu u_\mu\!-\!\nabla_\mu u_\nu$. 
Both $p$ and temperature $T$ are related to $\epsilon$ via
the QCD equation of state, for which we take the semi-realistic
result from Ref.~\cite{Laine:2006cp}.
%In Eq.(\ref{baseq}), $d_\alpha$ is the covariant derivative
%which is used to construct the time-like and space-like
%derivatives $D\equiv u^\alpha d_\alpha$ and 
%$\nabla_\mu\equiv d_\mu-u_\mu u^\alpha d_\alpha$, respectively.
%$\nabla_\mu\equiv \Delta_\mu^\alpha d_\alpha$, respectively.
%where
%$\Delta^{\mu \nu}=g^{\mu \nu}-u^\mu u^\nu$, respectively.
%The remaining definitions are 
%$\Delta^{\mu \nu}=g^{\mu \nu}-u^\mu u^\nu$ with the metric $g^{\mu \nu}$,
%$\langle\nabla^\mu u^\nu\rangle=
%\nabla^\mu u^\nu+\nabla^\nu u^\mu-\frac{2}{3} \Delta^{\mu \nu} \nabla_\alpha
%u^\alpha$ and the vorticity 
%$\omega_{\mu \nu}=\nabla_\nu u_\mu-\nabla_\mu u_\nu$. 
%For brevity, we refer to Ref.\cite{} for an explanation of the 
%remaining notation in Eq.(\ref{baseq}).
%
%
If the relaxation time $\tau_\Pi$ is not too small,
Eq.~(\ref{baseq}) are the most general shear viscous hydrodynamic
equations that are causal and guarantee that entropy
can never locally decrease \cite{oldpap}. 
%Note that 
%the terms in the last line of Eq.~(\ref{baseq}) 
Formally, Eq.~(\ref{baseq}) correspond to the relativistic
Navier-Stokes equations in the limit $\tau_\Pi\rightarrow 0$,
but contain corrections of higher order in gradients
for $\tau_{\Pi}>0$.

Unfortunately, the initial conditions for a hydrodynamic
description of an ultra-relativistic heavy-ion collision
at RHIC are poorly known, so one has to resort to model studies.
In order to describe Au+Au collisions at RHIC energies,
one typically assumes the energy density along the longitudinal
direction (the beam-line) to be ``boost-invariant'' to first 
approximation \cite{Bjorken:1982qr}. 
With this assumption, one still has to specify
the energy density distribution in the plane orthogonal to the
beam line (the transverse plane). At present, there exist
two main classes of models for this distribution, 
which we will refer to as Glauber-type
and Color-Glass-Condensate (CGC)-type models.
In the following, only Glauber-type models will be used.

\begin{figure}
\vspace{0.6cm}
\begin{center}
%\includegraphics[width=0.48\linewidth]{flucs.eps}
%\hspace*{0.25cm}
\includegraphics[height=5cm]{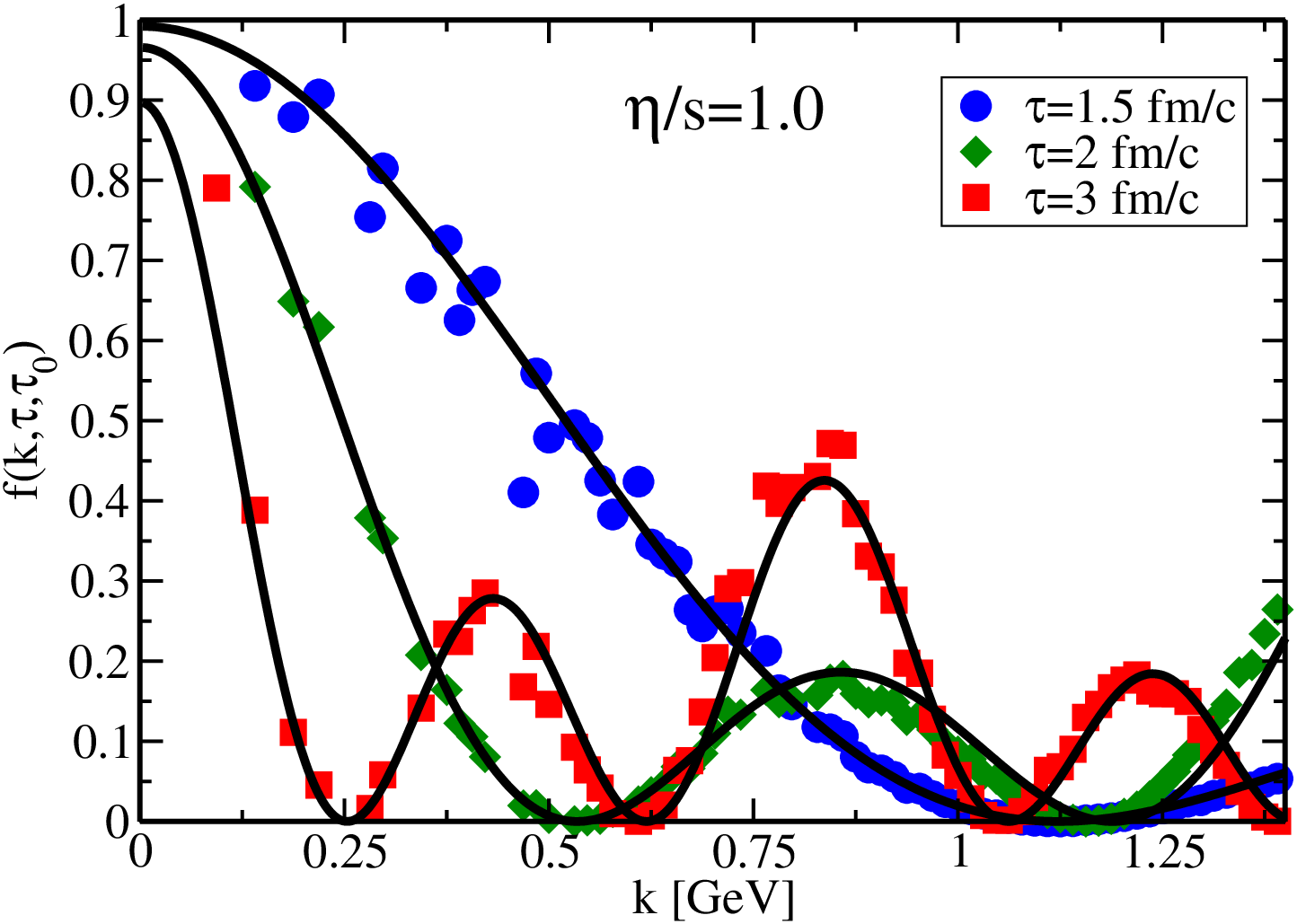}
\end{center}
\caption{Correlation function $f(k,\tau,\tau_0=1 {\rm fm/c})$ as a function
of momentum $k$, measured for our hydrodynamics code on a $64^2$ lattice with
a lattice spacing of $1 {\rm GeV}^{-1}$ (symbols), compared
to the ``analytic'' result from the linearized Eq.~(\ref{baseq}) (full lines).
The good overall agreement indicates the code is solving Eq.~(\ref{baseq})
correctly in the linear regime (see \cite{Baier:2006gy} for details).}
\label{fig:flucs}
\vspace*{-0.5cm}
\end{figure}

%The Glauber-type models build upon the nuclear thickness function
%$T_A({\bf x_\perp})$, defined as an integral 
%over the Woods-Saxon profile 
%$$
%T_A({\bf x_\perp})\!=\!\int_{-\infty}^{\infty}\!\!\!\! dz \frac{\rho_0}{
%1\!+\!\exp{[(\sqrt{{\bf x_\perp}^2+z^2}-R_0)/\chi]}},
%$$
%where $\rho_0$ is such that $\int d^2{\bf x}_\perp T_A({\bf x}_\perp)=A$
%with $A=197$, $R_0=6.4$ fm/c and $\chi=0.54$ fm/c for a gold nucleus.
%Using $T_A$, one can define the (two-dimensional)
%density of nucleons participating
%in the collision $n_{\rm Part}$ and that of binary collisions, $n_{\rm Coll}$
%as
%$$
%n_{\rm Part}({\bf x_\perp},{\bf b})=2 T_A({\rm x_\perp}+{\bf b}/2)
%\left[
%$$
The Glauber-type models build upon the Woods-Saxon density distribution
for nuclei, $\rho_A({\bf x})\!\sim\! 1/[1\!+\!\exp((|{\bf x}|\!-\!R_0)/\chi)]$,
where for a gold nucleus ($A\!=\!197$) we use $R_0\!=\!6.4$ fm/c, $\chi\!=\!0.54$ fm/c.
Integrating the Woods-Saxon distribution over the longitudinal direction
(corresponding to a Lorentz-contracted gold nucleus), one
obtains the nuclear thickness function $T_A({\bf x}_\perp)$. 
Contracting two $T_A$'s for the collision of two gold nuclei at 
a certain impact parameter ${\bf b}$, one 
can define number densities in the transverse plane,
such as the density of  participating
nuclei $n_{\rm Part}({\bf x_\perp},{\bf b})$ and the
density of binary collisions
$n_{\rm Coll}({\bf x_\perp},{\bf b})$ (see e.g. Ref.~\cite{Kolb:2001qz}).
As an initial condition for hydrodynamics, one then assumes
the energy density distribution
$\epsilon$ 
in the transverse plane to be proportional to either $n_{\rm Part}$
or $n_{\rm Coll}$ or a mixture of the two.  % \cite{Kolb:2001qz}.
In what follows, we will concentrate on the case
$\epsilon\sim n_{\rm Coll}$, since
%and $s \sim n_{\rm part}$, since 
for ideal hydrodynamics this provides a rough description
of the %${\bf b}$ 
centrality (or impact parameter) dependence of the 
total number of particles (``the multiplicity'') measured at 
RHIC \cite{Kolb:2001qz}.
%
%of RHIC multiplicity for different $b$ \cite{Kolb:2001qz}. 
Finally, for VH %viscous hydrodynamics 
one also has to provide
an initial condition for $\Pi^{\mu \nu}$. We 
choose the ``minimalist assumption'' $\Pi^{\mu \nu}\!=\!0$.
While one realistically expects $\Pi^{\mu \nu}$ to be
nonzero initially, this assumption translates to reducing
the effect of viscosity, which can serve as a baseline for future studies.

Because of boost-invariance, it is useful to
work in the coordinates $\tau\!=\!\sqrt{t^2\!-\!z^2}$
and $\eta\!=\!{\rm arctanh}(z/t)$ rather than $t,z$.
In these coordinates, boost-invariance dictates $u^\eta\!=\!0$,
so because of $u_\mu u^\mu\!=\!1$, the only non-trivial
fluid velocities can be chosen as $u^x,u^y$,
which are assumed to vanish initially.

Before discussing results from the numerics, 
one can get some intuition of viscous effects
on experimental observables by 
imagining the system to have a friction force
proportional to velocity.
In a heavy-ion collision, the expansion (at least initially)
is strongest along the beam axis, therefore one
expects viscosity to counteract this expansion.
In $\tau,\eta$ coordinates this is achieved  
by a reduction of the effective longitudinal pressure $p-\Pi^\eta_\eta$
through $\Pi^\eta_\eta\!>\!0$.
%from $\Pi^{\eta \eta}$. Since we assumed $T^\mu_\mu=0$
%and the equation of state is unchanged, 
Since initially $\Pi^{\tau}_\tau\!\ll\! \Pi^\eta_\eta$ but 
$\Pi^\mu_\mu\!=\!0$, the difference between equilibrium pressure $p$
and effective longitudinal pressure
has to appear as excess pressure in the transverse plane.
Therefore, viscosity should lead to higher transverse
velocities (``radial flow'') as compared to ideal hydrodynamics,
which is indeed the case \cite{Chaudhuri:2005ea,Baier:2006gy}.
Similarly, one can get an intuition of viscosity on
elliptic flow $v_2$ (the main angular modulation of radial flow for
non-central collisions): having a stronger reduction
effect on higher velocities,
%
%by reducing higher velocities more
%than lower ones,
%
%velocity differences,
viscosity tends to decrease velocity differences and hence 
elliptic flow. This agrees with
the qualitative trend found by Teaney \cite{Teaney}.

\begin{figure}
\vspace{0.6cm}
\begin{center}
%\includegraphics[width=0.48\linewidth]{flucs.eps}
%\hspace*{0.25cm}
%\includegraphics[height=5cm]{mult.eps}
\includegraphics[width=\linewidth]{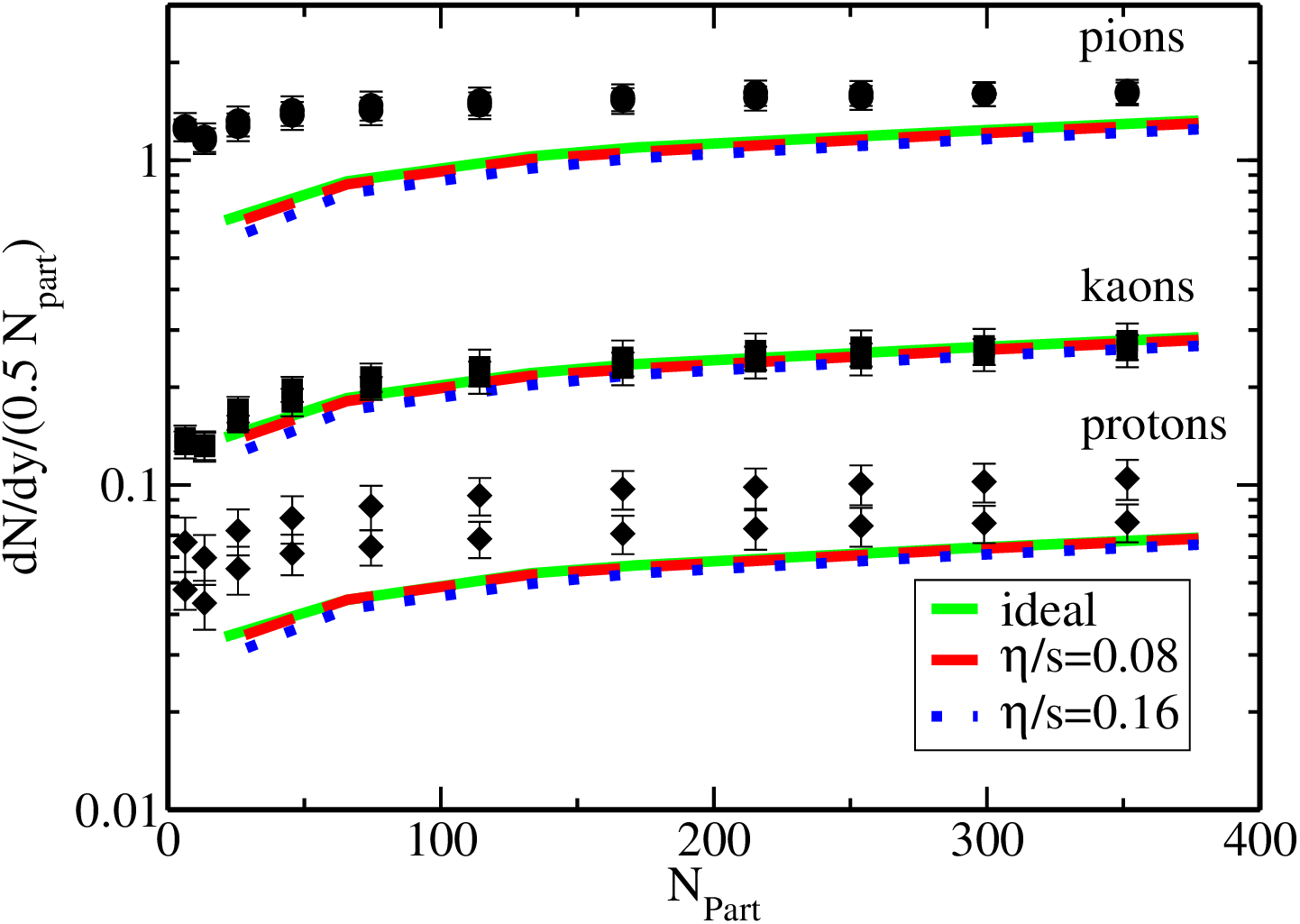}
\quad\newline
\quad\newline
\vfill
%\vspace*{2cm}
%\includegraphics[height=5cm]{meanpt.eps}
\includegraphics[width=\linewidth]{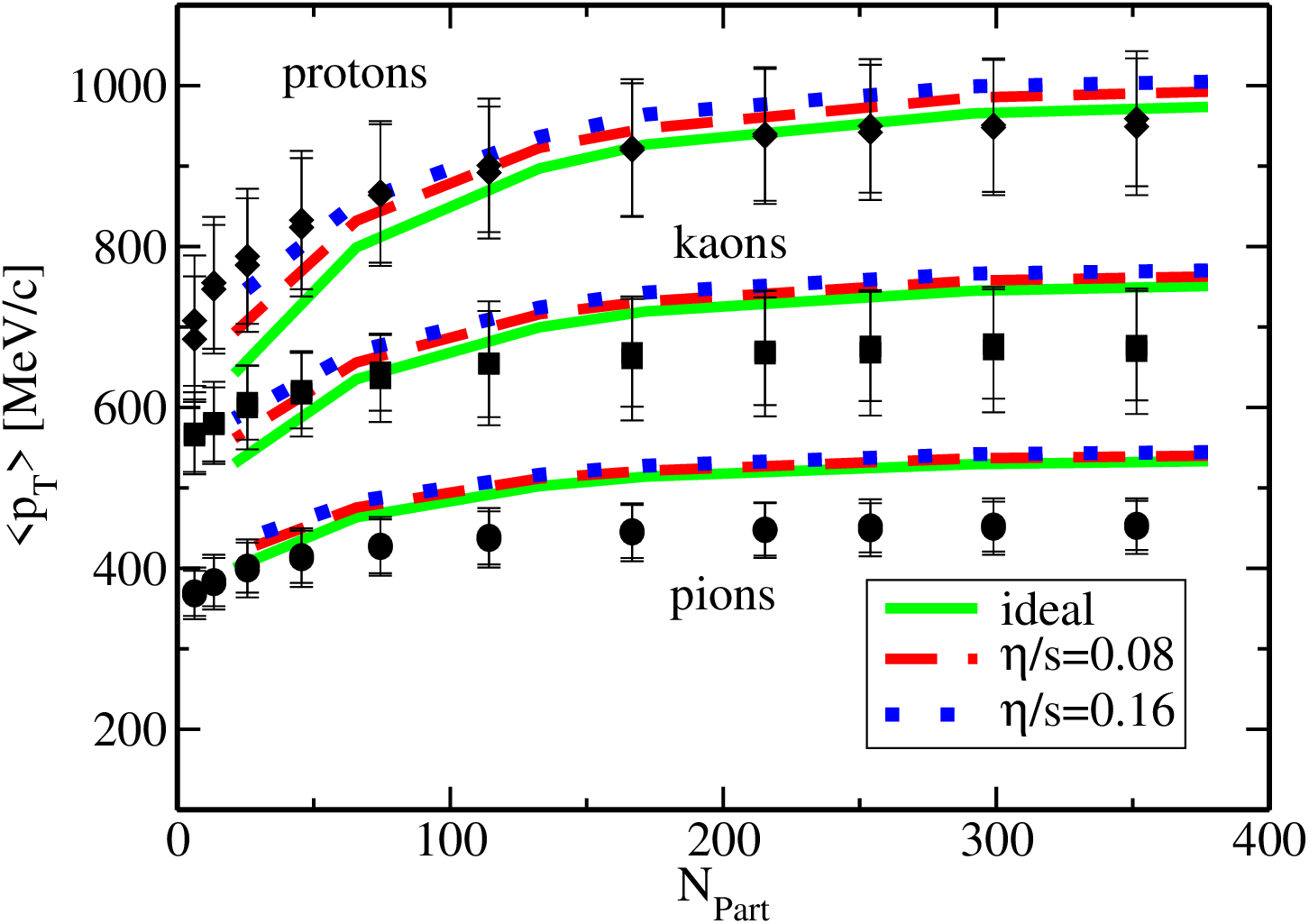}
\end{center}
\caption{Total multiplicity $dN/dy$ and mean momentum 
for $\pi^+,\pi^-,K^+,K^-,p$ and $\bar{p}$ 
from PHENIX \cite{Adler:2003cb}
for Au+Au collisions at $\sqrt{s}=200$ GeV, compared to 
our hydrodynamic model for various viscosity ratios $\eta/s$.}
\label{fig:mult}
\vspace*{-0.5cm}
\end{figure}

To solve Eq.~(\ref{baseq}) and treat the freeze-out (see below), we have used
a two-dimensional generalization
of the algorithm outlined in Ref.~\cite{Baier:2006gy}.
Details of the calculation will be given elsewhere \cite{inprep}.
We have checked that our algorithm agrees with the results
from Ref.~\cite{Romatschke:2007jx} for central collisions, when
dropping the extra terms in Eq.~(\ref{baseq}).
Also, our code passes the fluctuation test 
from Ref.~\cite{Baier:2006gy}, shown in Fig.~\ref{fig:flucs}.
%%
%
%performed a generalization of the fluctuation test 
%from Ref.\cite{Baier:2006gy}, shown in Fig.\ref{fig:flucs}.
%The agreement between measured and analytic results gives 
%us
%and found the 
%by R.Baier and one of us \cite{Baier:2006gy}. 
%The result for
%... are shown in Fig.bla.
We thus have some 
confidence that our numerical
algorithm solves Eq.~(\ref{baseq}) correctly.

When solving the set of equations (\ref{baseq}),
we set the ratio $\eta/s$ to be constant throughout the
evolution of the system, since modeling any
space-time dependence would necessarily introduce more unknown
parameters. Therefore, results on $\eta/s$ quoted below
should be considered as mean values over the entire
system evolution.

%\begin{figure}
%\vspace{0.6cm}
%\begin{center}
%%\includegraphics[width=0.48\linewidth]{flucs.eps}
%%\hspace*{0.25cm}
%\includegraphics[height=5cm]{meanpt.eps}
%\end{center}
%\caption{Total multiplicity for $\pi^+,\pi^-,K^+,K^-,p$ and $\bar{p}$
%from PHENIX \cite{Adler:2003cb}
%for Au+Au collisions at $\sqrt{s}=200$ GeV, compared to 
%our hydrodynamic model for various viscosity ratios $\eta/s$.}
%\label{fig:mpt}
%\end{figure}

To make contact with experiment, the hydrodynamic
variables are translated into particle spectra
via the Cooper-Frye freeze-out mechanism \cite{CooperFrye}
(adapted to VH \cite{Baier:2006um,Baier:2006gy}, see also \cite{Teaney}).
For simplicity, we use a single freeze-out
temperature $T_f$ but include 
the effect of resonance decays with masses up
to 2 GeV on the spectra \cite{Sollfrank:1990qz,Sollfrank:1991xm}.
The normalization of the initial energy
density and $T_f$ are chosen such that
the experimental data on total multiplicity and mean transverse
momentum $<p_T>$ as a function
of total number of participants 
$N_{\rm Part}=\int d^2{\bf x_\perp} n_{\rm Part}({\bf x_\perp},{\bf b})$
are reasonably reproduced by our model
(see Fig.~\ref{fig:mult}).
We choose to fit to kaons rather than pions
because the former are influenced less by 
Bose enhancement effects, 
which we have ignored \cite{Romatschke:2007jx}.
Note that for simplicity our model does not include a finite 
baryon chemical potential, prohibiting us to distinguish
particles from anti-particles. As a consequence, results
for protons cannot be expected to match experimental data.
Starting from ideal hydrodynamics with a freeze-out temperature
$T_f=150$ MeV, we have found that reasonable fits 
to $dN/dy$ and $<p_T>$ for VH %viscous hydrodynamics 
can be 
accomplished by keeping $T_f$ fixed and reducing the 
initial entropy density by $75\ \eta/s$ percent 
to correct for the viscous entropy production \cite{Romatschke:2007jx}.

\begin{figure}
\vspace{0.6cm}
\begin{center}
%\includegraphics[width=0.48\linewidth]{flucs.eps}
%\hspace*{0.25cm}
\includegraphics[width=\linewidth]{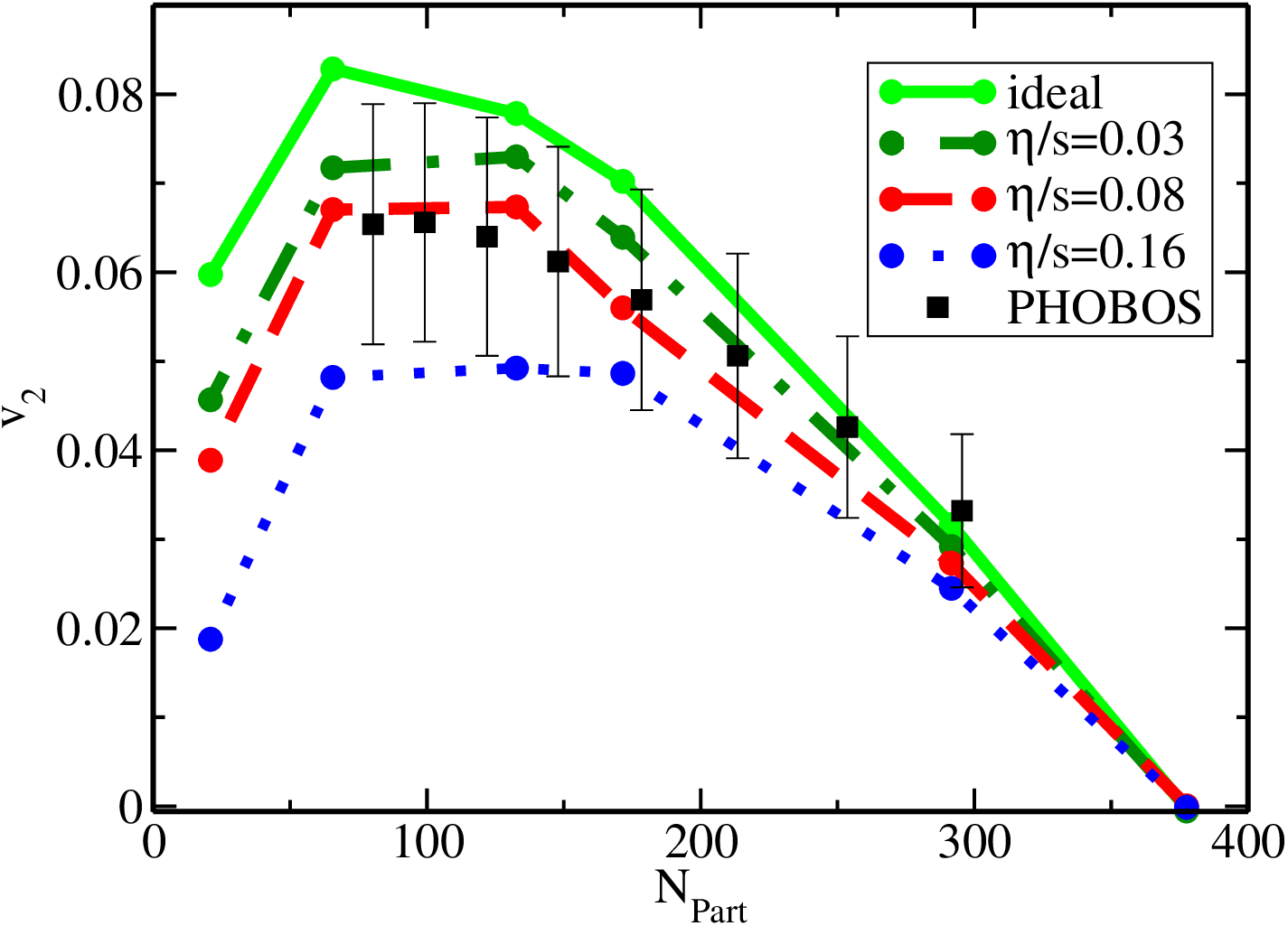}
\quad\newline
\quad\newline
\vfill
\includegraphics[width=\linewidth]{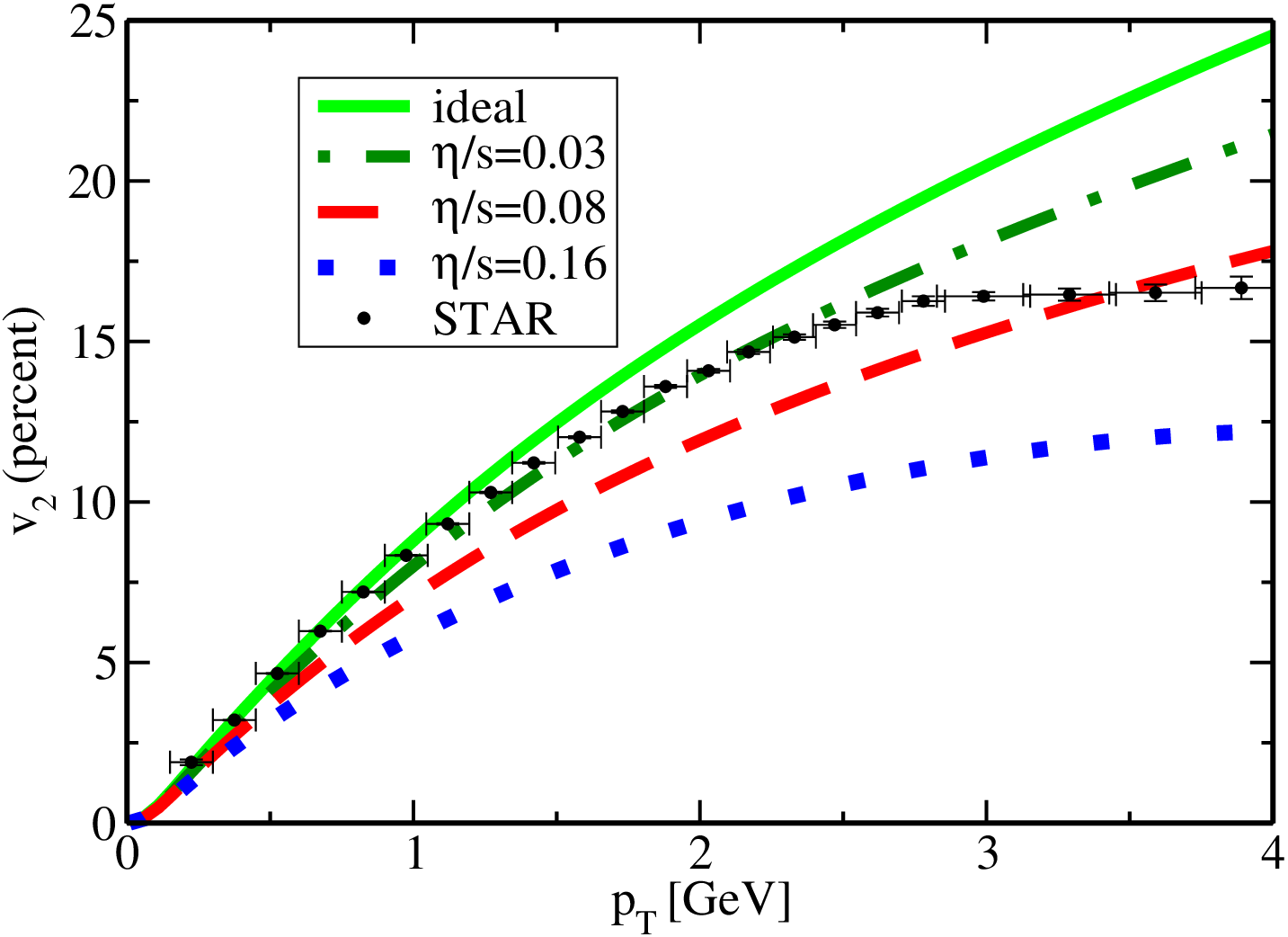}
\end{center}
\caption{PHOBOS \cite{Alver:2007qw} data on 
$p_T$ integrated $v_2$ and STAR \cite{Adams:2003zg} data on
minimum bias $v_2$, for charged particles in
Au+Au collisions at $\sqrt{s}=200$ GeV, compared to 
our hydrodynamic model for various viscosity ratios $\eta/s$.
Error bars for PHOBOS data show 90\% confidence level systematic errors
while for STAR only statistical errors are shown.}
\label{fig:v2}
\vspace*{-0.5cm}
\end{figure}

In Fig.~\ref{fig:v2} we compare our hydrodynamic model with
the above fit parameters to experimental data on 
the integrated and minimum bias
elliptic flow $v_2$, respectively.
%, integrated
%over $p_T$, and integrated over impact parameters (minimum bias), 
Shown are results for ideal hydrodynamics and VH %viscous hydrodynamics
for the initial condition $\epsilon \sim n_{\rm Coll}$ at an initial
time $\tau_0=1$ fm/c. The results hardly change 
when assuming instead $s \sim n_{\rm Part}$ as initial condition 
(see also \cite{Kolb:2001qz}) or varying $\tau_0$ by a factor of two. 
Interestingly, we also find that changing $\tau_\Pi$ hardly
affects the results shown. Note that this depends on the
presence of the terms in the last line of Eq.~(\ref{baseq}):
if these terms are dropped, increasing $\tau_\Pi$ tends to
further suppress $v_2$ in line with the trend found in 
\cite{Romatschke:2007jx}.

%with $\tau_\Pi=6 \eta/(\epsilon+p)$ and $\tau_\Pi\rightarrow0$ (extrapolated).
%The $\tau_\Pi\rightarrow0$ limit corresponds to an approximation
%linear in velocity gradients (first order), while for $\tau_\Pi>0$
%second order corrections are resummed. As can be seen from Fig.bla,
%these corrections %are comparatively small and 
%tend to reduce the effect from the first order approximation.
%Note that in the absence of
%the terms in the last line of Eq.(\ref{baseq}),
%this trend reverses and one recovers the trend found in 
%\cite{Romatschke:2007jx}.

For the above initial conditions, we have noted
that there is also hardly any effect from the
vorticity term. This can be understood as
follows: noting that for $u^\eta=0$ the 
only non-trivial vorticity is $\omega^{x y}$,
which vanishes initially because of $u^x=u^y=0$ and 
forming the combination $\nabla^x D u^y-\nabla^y D u^x$ we
find --up to third order corrections--
\beq
D \omega^{xy}+ \omega^{xy} \left[ \nabla_\mu u^\mu + \frac{D p}{\epsilon+p}
-\frac{D u^\tau}{u^\tau}\right]=\mathcal{O}(\Pi^3).
\label{vorteq}
\eeq
%for an equation of state with constant speed of sound.
This is the relativistic generalization of
the vorticity equation, well known
in atmospheric sciences \cite{Holton}.
Starting from $\omega^{xy}=0$,
Eq.~(\ref{vorteq}) implies a very slow buildup of
vorticity, explaining the tiny overall effect 
of the vorticity term in Eq.~(\ref{baseq}).
Note that upon dropping the %boost-invariance
assumption $u^\eta=0$, this term can become important
\cite{Rezania:1999gn}.

From Fig.~\ref{fig:v2} it can be seen that the effect
from viscosity on the elliptic flow is strong,
in line with estimates from Ref.~\cite{Teaney}.
Data on integrated $v_2$ is fairly well reproduced by
a viscosity of $\eta/s\sim 0.08$ and 
-- within systematic errors -- seems to be consistent
with $\eta/s\sim 0.16$. These values agree with
recent estimates by other groups 
\cite{Gavin:2006xd,Lacey:2006bc,Drescher:2007cd} and a lattice
QCD calculation \cite{Meyer:2007ic}.
However, the comparison to data for minimum bias $v_2$ in 
Fig.~\ref{fig:v2}
suggests that the ratio of $\eta/s$ is actually smaller
than the conjectured minimal bound $\eta/s=\frac{1}{4\pi}\simeq 0.08$ %obtained
%from the AdS/CFT duality.
As mentioned, this seems to be independent
from whether one adopts
$\tau_\Pi=6\ \eta/(\epsilon+p)$, the weak-coupling QCD result,
or extrapolates to $\tau_\Pi\rightarrow 0$, which is very close
to the AdS/CFT value found in \cite{Heller:2007qt}.
Indeed, the minimum bias $v_2$ seems to favor
$\eta/s\simeq0.03$, at least at low momenta, where
hydrodynamics is supposed to be most applicable.
Note that this result could change drastically 
if the minimum bias data were decreased by $20\%$, which
is the estimated systematic error quoted in \cite{Adams:2003zg}.
%(independent 
%
%(for 
%$\tau_\Pi=6 \eta/(\epsilon+p)$, the weak-coupling QCD result) 
%and $\eta/s=low$ 
%(for extrapolated $\tau_\Pi\rightarrow 0$, which is very close
%to the AdS/CfT value found in \cite{Heller:2007qt}).

There are, however, a number of caveats that should
be considered before taking the above numbers literally.
Firstly, we have only considered Glauber-type initial conditions,
and assumed $\Pi^{\mu \nu}(\tau_0)=0$.
It has been suggested that CGC-type initial conditions
lead to larger overall $v_2$ \cite{Hirano:2005xf}
which in turn would raise the allowed
values for $\eta/s$ in our calculation. This is due
to the larger eccentricities in this model \cite{Drescher:2006pi}
(note the issues raised in \cite{Lappi:2006xc}).
However, larger eccentricities in general also
lead to a faster build-up of transverse flow, which
is further enhanced by viscosity. Thus, when required to
fit all the data in Figs.~\ref{fig:mult} and \ref{fig:v2},
it is unclear whether this CGC-type model will
predict substantially higher $\eta/s$ than found here.
%in this work.
%what we found here.

Secondly, we used VH until the last scattering instead
of more sophisticated hydro+cascade models
(e.g. \cite{Bass:2000ib,Hirano:2005wx}). We do expect
changes in the extracted values of $\eta/s$ once 
a VH+cascade model description becomes available.
%we did not include hadronic rescatterings after
%freeze-out. These can be taken into account using 
%microscopic transport models 
%and do expect the cascade 
%inclusion of these 
%to affect our extracted values of $\eta/s$.
Finally, at present
we cannot exclude that effects not captured by hydrodynamics,
such as strong mean-fields, distort our results.
Work on QCD plasma instabilities and CGC dynamics might shed some light
on this issue.
% \cite{Romatschke:2006wg,Dumitru:2006pz,Arnold:2007cg}.

To summarize, we have presented the first viscous hydrodynamic
fits to experimental data on the centrality dependence of
$dN/dy$, $<p_T>$ and $v_2$ 
at top RHIC energies. For Glauber-type initial
conditions, we found that data seems to favor values 
for $\eta/s$ which are very small,
below the AdS/CFT bound \footnote{
While this work was being finalized, we became aware of
similar findings by another group \cite{Heinz}.}. While suggested to be possible
in \cite{Cohen:2007qr,Lublinsky:2007mm}, 
%recent estimates 
%\cite{Lacey:2006bc,Drescher:2007cd} and a lattice
%QCD calculation \cite{Meyer:2007ic} obey this bound.
%We thus see our number as a guideline,
it will be interesting to see whether 
%inasfar 
the above caveats -- once addressed -- can change our
results enough to accommodate viscosity equal or larger 
than 
the bound.
%
%
%%
%
%addressing the above caveats is needed to ultimately
%decide on the validity of this bound.
%
%drawing firm conclusions.
%
%to address the above caveats.
%
%The prudent way to interpret our numbers is thus as a guideline,
%with more work needed to address the above caveats.
In any case, we hope that our work can serve as a guideline
to understanding the properties of the fluid created at RHIC.
%to unraveling the mysteries of the fluid created at RHIC.
%
%to understanding this interesting physics offered by RHIC.
%
% this interesting physics at RHIC.

\acknowledgments

PR would like to thank P.~Huovinen and T.~Lappi for fruitful discussions.
%, M.~Laine,
%G.A.~Miller and S.~Salur for fruitful discussions. Special thanks go to
%P.~Huovinen for providing me with the data file for the resonance
%decays and to M.~Laine for providing the tabulated equation of state.
This work was partially supported by the US Department of Energy, grant
number DE-FG02-00ER41132.

\end{document}